\documentclass[pre,preprintnumbers,amsmath,amssymb,aps,showkeys,preprint,showpacs]{revtex4-1}

\usepackage{graphicx}% Include figure files
\usepackage{dcolumn}% Align table columns on decimal point
\usepackage{bm}% bold math
\usepackage{xspace}

\begin{document}

%\preprint{APS/123-QED}

\title{The Stability of Boolean Networks with Generalized Canalizing Rules}% Force line breaks with \\

\author{Andrew Pomerance}
 \email{pomeranc@umd.edu}
\author{Michelle Girvan}
\author{Ed Ott}
\affiliation{%
Institute for Research in Electronics and Applied Physics \\
University of Maryland, College Park, MD, 20752
}%

\date{\today}% It is always \today, today,
\begin{abstract}
Boolean networks are discrete dynamical systems in which the state (zero or one) of each node is updated at each time $t$ to a state determined by the states at time $t-1$ of those nodes that have links to it.  When these systems are used to model genetic control, the case of 'canalizing' update rules is of particular interest. A canalizing rule is one for which a node state at time $t$ is determined by the state at time $t-1$ of a single one of its inputs when that inputting node is in its canalizing state.  Previous work on the order/disorder transition in Boolean networks considered complex, non-random network topology.  In the current paper we extend this previous work to account for canalizing behavior.
\end{abstract}

%Use showkeys class option if keyword
\maketitle

\newcommand{\lR}{\ensuremath{\lambda_R} \xspace}
\newcommand{\sk}{\sigma_{\mathcal{K}(i)}}
\newcommand{\skh}{\hat{\sigma}_{\mathcal{K}(i)}}
\newcommand{\Kin}[1][i]{\ensuremath{{K}^{in}_{#1}} \xspace}
\newcommand{\genp}[3]{\ensuremath{p^{(#2, #3)}_{#1}} \xspace}
\newcommand{\gps}[1]{\genp{i}{k}{#1}}
\newcommand{\gp}{\genp{i}{k}{s}}
\newcommand{\ps}[1][i]{\ensuremath{p^*_{#1}} \xspace}
\newcommand{\kset}[1][i]{\ensuremath{\mathcal{J}_i} \xspace}

\section{\label{sec:intro}Introduction}

Boolean networks have been extensively studied as models for genetic control of cells \cite{Kauffman1969, Kauffman1993}.  In this framework, the genetic regulatory network is modeled as a directed graph, where links correspond to the influence of one gene on the expression of another.  Individual genes are either off or on, represented as 0 or 1, respectively, and the state of a gene at time $t+1$ is given by a Boolean update function of the states of its inputs at time $t$.  In early analyses, both the network topology and the update functions were randomly assigned.  In particular, Kauffman's $N-K$ network model \cite{Kauffman1969, Kadanoff} has received significant study.  According to this model, there are $N$ nodes (genes) in the network, each having the same number of input links, $K$, and the nodes from which these input links originate are chosen randomly with uniform probability.  Additionally, the update function determining the time evolution at each node is defined by a random, time-independent, $2^K$-entry truth table that can be characterized \cite{Derrida1986} by the `bias' $p$, which, as discussed subsequently, is the probability that a one appears as the output of the update function.  Using the Hamming distance between two states of the system (i.e., the number of nodes for which the states disagree) as the distance measure, these systems, when large, exhibit both disordered (unstable) behavior, where the distance between typical initially close states on average grows in time, as well as ordered (stable) behavior, where the distance decreases.  Separating the two is a `critical' regime.  

So-called `canalizing functions' are a significant modification of the random truth table model of previous work.  Canalizing functions, believed to be relevant to genetic networks \cite{PNAS1, harris}, are those functions where an argument of the function (the `canalizing input'), having a certain value (the `canalizing value'), determines the value of the function independently of the values of the other arguments of the function (inputs) \cite{Kauffman1993}.  If the canalizing input does not have the canalizing value, the function is determined by the other inputs.  (Further refinements can include a hierarchy of canalization \cite{PNAS2}.)  Previous work \cite{PNAS2, chaos, drossel, socolar} has considered random network topology and random choices of canalizing inputs and, with those assumptions, has demonstrated that canalizing functions often stabilize networks that would be disordered in their absence.  A key quantity, defined in previous work by Shmulevich and Kauffman, is the `activity' \cite{activities, activities2} of a Boolean variable on a Boolean function.  This quantity can be used to characterize the increased importance of canalizing inputs and plays a crucial role in the theory presented in this paper.

%According to Eq. \eqref{eq:qstability}, the predicted response of a gene network to silencing a gene is increased stability.  To see this, we note that $\lambda_Q$ is monotonic in the $q_i$, and silencing a gene is equivalent to setting that node's sensitivity to zero.  This results in a smaller $\lambda_Q$, which implies increased stability of the network.  However, ref. \cite{123} reports that a mutant strain of macrophage that has a gene silenced actually exhibits mildly chaotic behavior, in contrast to the unmutated case which exhibits criticality, which is inconsistent with the theory of ref. \cite{self}.  This source of this disagreement is the presence of canalizing inputs, which will be demonstrated below.

%Boolean networks have also been used as models of neural networks \cite{hopfield}.  In the Boolean model, each neuron is either activated or quiescent, and neurons activate one another through their connections.  A given neuron becomes activated at a given timestep if a weighted sum of its inputs exceeds a threshold.  These rules will be canalizing if a single input has a significantly larger weight than the rest of the inputs; in general, however, these networks will be `quasicanalizing,' a situation where a single node or a small group of nodes change the \textit{probability} of activation/quiescence independent of other nodes, but are not the sole determination.  This concept will be discussed in depth in Secs. III and VI.

An important hypothesis in genetic networks is that biological systems exist at the transition between order and disorder \cite{Kauffman1969} (i.e., the `life at the edge of chaos' hypothesis).  It has also been suggested that this transition may be relevant to the onset of cancer \cite{self}.  This paper is concerned with deriving a condition determining the `edge of chaos' in Boolean networks taking into account the specific topology of the network (as was treated in Ref. \cite{self}) and canalizing update rules.  In Sec. II, we present definitions that will be used in the remainder of the paper and summarize previous results on Boolean network stability.  In Sec. III, we present a generalized probabilistic model of canalizing behavior, and we use this model with the Shmulevich-Kauffman activity to extend the results of Ref. \cite{self} to the case of networks with canalizing functions.  We derive a stability criterion that offers an advancement over previous work on Boolean network models with canalizing rules in two ways: (1) it accounts for the specific topological properties of a considered Boolean networks, rather than purely random topology; and (2) it accounts for possible correlation between network topology and canalizing behavior, e.g., correlation between the choice of canalizing input and local topology.  In Sec. IV, we numerically test our derived stability criterion on a variety of complex network topologies, and in Sec. V we explore the effect of correlation between strict canalization and network topology.

\section{Preliminary Definitions and Previous Results}
A Boolean network is defined by a state vector $\Sigma(t) = [\sigma_1(t) \sigma_2(t) ... \sigma_N(t)]^T$, where each $\sigma_i \in \{0, 1\}$, and a set of update functions $f_i$, such that
\begin{equation}
\label{eq:definition}
\sigma_i(t) = f_i(\sigma_{j(i,1)}(t-1), \sigma_{j(i,2)}(t-1), ... ),
\end{equation}
where $j(i,1), j(i,2),..., j(i, K^{in}_i)$ denote the indices of the $K^{in}_i$ nodes that input to node $i$, and we denote this set of nodes by $\kset = \{j(i,k)| k=1,2,..,K^{in}_i\}$.  [In the following discussion, $k$, which is between 1 and $K_i^{in}$, is used to label an input to node $i$, or, equivalently an argument of $f_i$; $j$, which is between 1 and $N$, refers to the network index of the node corresponding to input $k$; $k(i,j)$ and $j(i,k)$ are used to switch between them.  Similarly, $\sigma_j$ is the state of node $j$, and $s_k$ is the $k$-th input to $f_i$.]  The number of input links $K^{in}_i$ to node $i$ is called its in-degree, and the number of output links $K^{out}_i$ from node $i$ is called its out-degree.  The update function $f_i$ at each node $i$ is usually defined by a truth table, where the table for node $i$ has $2^{K^{in}_i}$ rows, one for each possible set of the states of the $K^{in}_i$ nodes that input to node $i$, and each input state row is followed by its resulting update output state for node $i$, thus forming a $2^{K^{in}_i}$ entry output column.  Thus the $K_i^{in}$ input entries of the $2^{K^{in}_i}$ rows of the truth table represent all possible arguments of $f_i$ in Eq. \eqref{eq:definition}, and the output column gives the value that $f_i$ assumes for that set of its arguments.

An important property of a given Boolean network is whether it is `ordered' or `disordered.'  This property is defined in large networks by considering the trajectories resulting from two close initial states, $\Sigma(t)$ and $\tilde{\Sigma}(t)$. To quantify their divergence, the Hamming distance of coding theory is used: $h(t) = \sum_{i=1}^N |\sigma_i(t) - \tilde{\sigma_i}(t)|$, and the two initial conditions are `close' if $h(0) << N$.  If the network is ordered, on average $h(t) \rightarrow 0$ as $t \rightarrow \infty$.  In disordered networks, $h(t)$ quickly increases to $O(N)$, while a `critical' network is at the border separating the two regimes.  (In this paper, we use the word `stable' to refer to ordered networks.)

To obtain an analytically tractable model for the stability of $N-K$ Boolean networks, Derrida and Pomeau \cite{Derrida1986} considered an `annealing' procedure applied to the original problem and calculated the probability that, after $t$ steps, a node state is the same on two trajectories that originated from initially close conditions.  Annealing, in this context, refers to assuming that both the $f_i$ at each node and the network of connections are randomly reassigned at each time step, in contrast to the `quenched' problem, in which both the update functions and topology are constant for all time.  Later authors generalized the Derrida-Pomeau analysis to include variable in-degree \cite{Luque1995, Luque1997, Fox2001} and joint in-degree/out-degree distributions \cite{Lee2008}.  In the annealed situation, at each time step $t$ \textit{both} the truth table outputs \textit{and} the network of connections are randomly chosen.  The actual situation of interest, however, is the case of `frozen-in' networks, where the truth table and network of connections are fixed in time.  It was hypothesized and subsequently numerically confirmed that, for large networks, results obtained in the analytically tractable annealed situation are the same as those in the analytically intractable frozen situation; this is referred to as the `annealing approximation.'

In Ref. \cite{self}, we developed a more general approximate technique for determining the stability of large Boolean networks.  This technique used what we called a `semiannealing' procedure, which differed from the situation considered by Derrida in that \textit{only} the truth table outputs are randomly reassigned at each time step; the network of connections was kept frozen.  By using this procedure, one can investigate the effect of given specific network topology on the stability of a Boolean network.  Numerical experiments confirmed the predictions of the semiannealed theory \cite{self} in cases of complex network topology, exploring such effects as correlation between the number of inputs and outputs at each node \cite{eigenvalue}, assortativity \cite{Newman2002}, community structure \cite{Girvan}, and the presence of small motifs similar to those found in biological systems \cite{alon}.  These cases cannot be accounted for using the full annealing approximation.

When considering annealed Boolean truth tables, the update function of each node, $f_i$, is solely parameterized by the `bias' of node $i$, denoted $p_i$.  The bias is the probability that any given output entry in the truth table is a one.  This can be recast in terms of the `sensitivity' of node $i$, denoted $q_i$, which is the probability that differing inputs to $f_i$ yield a differing output of $f_i$.  In the annealed truth table case, $q_i = 2p_i(1-p_i)$.  

However, this parameterization is not complete for canalizing functions.  In Ref. \cite{drossel}, Kaufman et al. consider the case of $N-K$ networks with $K=2$ and non-annealed truth tables.  For $K=2$ networks, the possible Boolean functions fall into four classes: frozen functions (i.e., those functions that output the same value, one or zero, independent of their inputs), non-canalizing functions, and two types of canalizing functions.  Reference \cite{drossel} shows that, when there are fewer nodes with non-canalizing functions than with frozen functions (by the addition of either frozen functions or canalizing functions), the network is in the ordered regime.  Another approach to analyzing stability of Boolean networks with canalizing functions is to consider `damage spreading' (i.e., the propagation of disagreement between $\sigma_i$ and $\tilde{\sigma}_i$).  This approach was used in Ref. \cite{socolar} for Boolean networks with canalizing rules, with results similar to \cite{drossel}.

In Sec. III, we refine the semiannealed procedure of Ref. \cite{self}, which relies heavily on the sensitivity, to account for canalizing behavior.  However, as discussed above, the sensitivity does not include sufficient information to account for canalizing behavior.  This quantity treats all input nodes equally, which is clearly not appropriate for canalizing functions.  Instead, we make use of the `activity' of node $j$ on node $i$, $r_{ij}$ (or, equivalently, the activity of input argument $k(i,j)$ on update function $f_i$, $\hat{r}_{ik}$).  First described by Shmulevich and Kaufmann in Ref. \cite{activities}, the `activity' of input $k$ on a Boolean function $f_i$ is the probability that a bit flip of \textit{only} input $k$ leads to a flip of the output of $f_i$ if all inputs other than $k$ are independent and have equal probabilities of being one or zero.  This definition clearly emphasizes canalizing inputs over non-canalizing inputs \cite{activities}.  In our presentation of generalized canalization below, we calculate the activities for truth tables with this property and use the probabilistic interpretation of our definition to define a new semiannealing procedure, which has the activities as its defining parameters.

\section{Generalized Canalization and the Stability Criterion}

The semiannealing procedure used in Ref. \cite{self} independently and randomly reassigned the output elements of the truth table governing node $i$ to be one or zero with probability $p_i$ or $1-p_i$, respectively.  This is not true of canalizing functions: if the canalizing input takes its canalizing value in a given row of the table, the probability of a one appearing in the corresponding entry of the output column is zero (or one).  We call this behavior `strictly canalizing.'  We now introduce a generalization of strictly canalizing behavior to the case of `quasicanalizing' inputs, which we define as the case where the probability \gp that a one appears in the output of node $i$'s truth table if input $k$ takes value $s$, averaged over all other inputs, depends on $s$.  We illustrate this concept in Tables 1(a)-(c).  (The average of \gp, weighting the states $s_l = 1$ and $s_l = 0$ equally for all $l \neq k$, is $\ps = (\gps{0} + \gps{1})/2$, which we call the `truth table bias.'  This quantity will be discussed in detail below.)  Strict canalization with respect to input $k$, illustrated in Table 1(a), is the case when $\gp = 0$ or $1$ when $s$ is the canalizing value.  If $\gps{0} = \gps{1}$, $k$ is a non-canalizing input to $i$, and Table 1(b) consists entirely of non-canalizing inputs.  Quasicanalizing inputs are neither strictly canalizing nor non-canalizing: if $\gps{0} \neq \gps{1}$ and $0 < \gp < 1$ for $s = {0,1}$, then input $k$ is said to be quasicanalizing.  Table 1(c) illustrates this property.  We call truth tables where all inputs are non-canalizing `unstructured' (e.g., Table 1(b)), and those with any canalizing inputs, strict or quasi-, `structured' (e.g., Tables 1(a) and (c)).   We note that a complete, realizable set of \gp is highly constrained, and the constraints are given in the Appendix.
\begin{table}
\begin{tabular}{|c|c|}
\hline
Inputs & $\Pr[f=1]$ \\ \hline
00&$p$\\
01&$p$\\
10&0\\
11&0\\
\hline
\multicolumn{2}{c}{(a)}
\end{tabular}
\begin{tabular}{|c|c|}
\hline
Inputs & $\Pr[f=1]$ \\ \hline
00&$p$\\
01&$p$\\
10&$p$\\
11&$p$ \\
\hline
\multicolumn{2}{c}{(b)}
\end{tabular}
\begin{tabular}{|c|c|}
\hline
Inputs & $\Pr[f=1]$ \\ \hline
00&$p_1$\\
01&$p_1$\\
10&$p_2$\\
11&$p_2$ \\
\hline
\multicolumn{2}{c}{(c)}
\end{tabular}
\caption{Tables illustrating three types of canalizing behavior in a function with two inputs.  In (a), the first input is strictly canalizing: when it is one, the output of the function is guaranteed to be zero; if it is zero, the probability that the function takes the value one is $p$.  In (b), neither input is canalizing and, for each input, the output of the function is independently one with probability $p$.  In (c), the first input is quasicanalizing: if it one, the function is one with probability $p_2$, otherwise it is one with probability $p_1 \neq p_2$.  In all three cases, the second input is noncanalizing.}
\end{table}

% include table here

We now write down the procedure by which we can make random draws of the truth tables conforming to the set of \gp that defines the update function $f_i$.  The details of the calculation are given in the Appendix, but we state here that, for a given draw the probability that $f_i$ is one, $Pr[f_i(s_1, s_2, ..., s_{K^{in}_i}) = 1] \equiv \phi_i(s_1, s_2, ...)$, is given by
\begin{equation}
\label{eq:probf}
\phi_i(s_1,... s_{\Kin}) = (p_i^*)^{1-\Kin}\prod_{k=1}^{\Kin}\genp{i}{k}{s_k}.
\end{equation}
This defines an ensemble of update functions for each node $i$, where the sampling probability of each member can be derived from \eqref{eq:probf}.  Loosely speaking, we would like to be able to average the dynamics of all networks over this ensemble to derive the stability criterion (i.e., the boundary between the stable and unstable phases in the space of all \gp).  However, this is analytically intractable since the dynamics of a given network depend crucially on the exact nodal update functions (e.g., frozen functions which are sampled with non-zero probability), and the vast number of sets of ${f_i}$ makes member-by-member calculation infeasible.  Instead, we use a semiannealing sampling procedure to approximate this average by considering a system where, at each time step, a different randomly chosen member of the ensemble of $f_i$ is used to advance the system.  Thus, at each time step, we randomly assign the truth table output of $f_i$, when presented with inputs $(s_1, s_2, ...)$ to be one with probability $\phi_i(s_1, s_2,....)$ and zero with probability $1-\phi(s_1,s_2,...)$.  The motivation for this semiannealed approach, as well as for other previous annealing approaches, is (i) that they are analytically tractable, and (ii) that it is supposed that, for large networks, the stability in the annealed situation approximates that of a typical (with respect to the ensemble) frozen-in case.  Here by the frozen-in case we refer to a system with a given temporally constant network topology and set of truth tables, which are randomly assigned initially using the probabilities in the annealing procedure. We numerically confirm that the stability of frozen-in networks corresponds well with that of our semiannealed networks (Sec. IV).

We now follow Refs. \cite{activities, activities2} and introduce a quantity $r_il$ that we call the `activity', which we define as the probability that, at any time on an orbit, the output of $f_i$ changes if input $l$ is flipped (either from 0 to 1 or vice versa), keeping the other $K^{in}_i-1$ inputs unchanged.  The activity $r_il$ will play an important role in the derivation of the stability criterion.  In order to obtain the activities, we first make the assumption that orbits on the semiannealed network are ergodic, and, given such an orbit, we can define $\rho_i$ as the fraction of time that the state $\sigma_i = 1$; we call this the `dynamic bias' since it is determined by the dynamics of the network, in contrast to the truth table bias above.  Assuming independence of the probability of nodal input states to node $i$ (appropriate to our supposition of locally-tree like topology, discussed below), we have 
\begin{equation}
\label{eq:dynbias}
\rho_i = \sum_{\{s_k\}_i} \phi_i(\{s_k\}_i) \rho(\{s_k\}_i),
\end{equation}
where the sum is over all possible $2^{K_{in}^i}$ inputs to node $i$, which is denoted $\{s_k\}_i$; $\rho(\{s_k\}_i)$ is the probability that the set of states of the input nodes to node $i$, $\{\sigma(t)_j\}_i$, takes the values in the set $\{s_k\}_i$ (where again $k$ denotes the argument number to the update function $f_i$, and $j$ refers to a node index in the network); and $\phi_i(\{s_k\}_i)$ is a shorthand notation for the quantity in Eq. \eqref{eq:probf}.  Note that $\phi_i(\{s_k\}_i)$ is a purely local quantity; i.e., it is determined from the random truth table annealing process for node $i$ and is independent of the truth table assignments at other nodes.  The probability of the inputs to node $i$ taking the values $\{s_k\}_i$ can in turn be written in terms of the dynamic biases of the input nodes,
\begin{equation}
\label{eq:dynbias2}
\rho(\{s_k\}_i) = \prod_{j\in \kset} [ s_{k(i,j)}\rho_j + (1-s_{k(i,j)})(1-\rho_j)],
\end{equation}
where $\kset = \{j | A_{ij} = 1\},$ and $k(i,j)$ is used to convert from node index $j$ to argument number $k$.  In principle, one could insert Eq. \eqref{eq:dynbias2} into Eq. \eqref{eq:dynbias} and iteratively solve for all $\rho_i$ (provided the iteration converged) or, alternatively, one could numerically generate a long orbit and approximate $\rho_i$ as the fraction of time that $\sigma_i = 1$ on the orbit.  In this paper, as an example, we will restrict our consideration to a case where a very simple exact solution to Eqs. \eqref{eq:dynbias} and \eqref{eq:dynbias2} is available, and we will base our numerical experiments in Sec. IV on that case.

We now define $\phi_i^{(l,s)} \equiv \phi_i(s_1,...,s_{l-1}, s, s_{l+1}, ...)$ to be a $\Kin-1$ input function that denotes the probability that the truth table output is one if input $k$ is $s$ given some $\Kin-1$ element set of other inputs; again, we emphasize that the $\phi_i^{(l,s)}$ are local quantities.  (The input to this function is a reduced set of inputs $\{s_{k}\}_i^l$, which is related to $\{s_{k}\}_i$ by removing the $l$-th element; we suppress the explicit dependence below.)  With this, we calculate the activity as
\begin{equation}
\label{eq:genactivity}
r_{ik} = \sum_{\{s_{k}\}_i^l} \big[ \phi_i^{(l,0)}(1-\phi_i^{(l,1)}) + \phi_i^{(l,1)}(1-\phi_i^{(l,0)}) \big ] \rho(\{s_{k}\}_i^l),
\end{equation}
where $\rho(\{s_{k}\}_i^l)$ is the probability of node $i$ having the reduced input set $\{s_{k}\}_i^l$.  In the numerical tests in Sec. IV, we consider this probability to be uniform over all $2^{\Kin-1}$ possibilities; i.e., we consider the trivial solution to Eqs. \eqref{eq:dynbias} and \eqref{eq:dynbias2} given by $\rho_i = 1/2$ and $\rho(\{s_{k}\}_i) = 2^{-K_{in}^i}$.  This implies that both canalized and canalizing values are 0 or 1 with probability one-half and, further, that truth table biases are symmetrically distributed around one-half.  However, in cases where there is a bias in the canalization, e.g., a node is more likely to be canalized to one, or truth table biases are not symmetrically distributed, $\rho(\{s_{k}\}_i)$ is no longer uniform, and a procedure to calculate the $\rho_i$ for each node as described above may be employed.  

Our semiannealing procedure defines an ensemble of nodal truth table time courses, and we can define probabilities of the dynamically evolving states for an arbitrary member of this ensemble.  In particular, as in Ref. \cite{self}, we define the $N$-dimensional vector $\vec{y}(t)$, where each element $y_i(t)$ tracks the probability that node $i$ differs between two initally close states after $t$ time-steps: $y_i(t) = \Pr[\sigma_i(t) \neq \tilde{\sigma}_i(t)]$.  Our goal is to find an update equation for $y_i(t)$ and perform linear stability analysis on the solution $y_i(t) = 0$ to obtain the stability criterion.  The update equation will be derived under the assumption that the inputs $y_j(t)$ are statistically independent of one another.  This assumption holds in the case of locally tree-like topology \cite{self, treelike}.

Since we are performing linear stability analysis, we can make several simplifying approximations.  The probability of $d$ inputs to node $i$ being different between the trajectories of two initially close states is of order $\mathcal{O}(y^d(1-y^{\Kin-d})) \approx \mathcal{O}(y^d)$.  Linear stability applies for $y$ small, so the case of multiple inputs to $f_i$ are flipped can be ignored, and we can approximate the probability that only input node $j$ to node $i$ is flipped as $y_j(t)$.  The probability that the single bit flip of node $j$ causes a flip in the output of $f_i$ is the activity $r_{ij}$.  Since the input flipping and the the probability that this leads to a flip in the output of $f_i$ are independent, the probability that a flip in node $j$ occurs and leads to a flip in the node $i$ in the next time step is thus $r_{ij}y_j(t)$.  In the linear stability limit, we can sum these probabilities inputs $j$ to get the following approximate evolution equation for small perturbations from the solution $\vec{y}(t) = 0$:
\begin{equation}
\label{eq:update}
y_i(t+1) \approx \sum_{k = 1}^{\Kin} r_{ik}y_{j(i,k)}(t) + \mathcal{O}(y^2).
\end{equation}
This can be written in matrix form after discarding the higher-order terms as
\begin{equation}
\vec{y}(t) \approx R \vec{y}(t-1), 
\end{equation}
where $R$ is the `activity matrix' with elements $R_{ij} = r_{ik}$ if there is a link from $j$ to $i$ ($k = k(i,j)$), and zero otherwise.  From this equation, we see that stability is determined by the largest eigenvalue $\lR$ of this matrix: 
\begin{eqnarray}
\nonumber \lR &>& 1, \text{$y = 0$ is unstable;} \\
\label{eq:stability} \lR &=& 1, \text{$y = 0$ is critical;} \\
\nonumber \lR &<& 1, \text{$y = 0$ is stable.}
\end{eqnarray}
Note that, since the elements of the matrix $R$ are all non-negative, the Perron-Frobenius theorem guarantees that $\lambda_R$ is real and non-negative.

Before concluding this section, we note the relationship of \eqref{eq:stability} to the case of unstructured truth tables \cite{self}.  When there are no canalizing inputs present, the right hand side of Eq. \eqref{eq:probf} reduces to $\phi_i = \ps = p_i$ for all possible inputs, and $r_{ik} = 2p_i(1-p_i) \equiv q_i$ is constant across each row of the $R$ matrix.  This is the central result of Ref. \cite{self}.

\section{Numerical Results}

In this section, we present numerical results testing our derived criterion for the stability of Boolean networks with canalizing truth tables.  We test the theory by measuring the long-time Hamming distance between trajectories that differ in only a few initial states as a function of $\lambda_R$.  We vary $\lambda_R$ in two ways: (a) a varying proportion of nodes have a single, strictly canalizing input; and (b) each node has a single quasicanalizing input, and that input has varying activity on each node's function.   In Sec. IV.A, we treat case (a) with several network degree distributions.  We consider $N-K$ networks, which is most comparable to previous work on canalizing inputs \cite{drossel, socolar}, as well as networks with power-law in-degree distribution.  We treated the cases of assortativity/disassortativity and community structure in Ref. \cite{self}, and we believe these cases can be treated in an analogous way using the activity matrix.  In Sec. IV.B, we consider case (b) with networks of the same general topologies.  The networks under consideration in Sec. IV.B, however, have more average inputs as compared to Sec. IV.A.

Our general approach to test the transition from order to disorder is the following.  We begin with a given network topology and, using the techniques of Ref. \cite{self}, calculate a uniform sensitivity $q$ for all nodes that would yield a Boolean network slightly in the disordered regime.  We then choose a uniform truth table bias $\ps = p^*$  for all nodes that satisfies $q = 2p^*(1-p^*).$  Thus, for all networks under consideration in this section, Ref. \cite{self} would predict that the network is slightly disordered.  We then vary $\lambda_R$ by methods (a) or (b) above.  As discussed in Sec. V, the addition of canalizing behavior tends to stabilize a network, and thus the transition is approached from the disordered regime.  We choose the canalizing input of each update function randomly.

Figures 1 and 2 demonstrate the results of our numerical tests.  Each data point in Figs. 1 and 2 is the average steady-state Hamming distance measurement of 1000 different frozen realizations of the truth tables.  Each network has $N = 10^5$ nodes, and the steady-state Hamming distance is calculated as the average Hamming distance from time $t=900$ to $t = 1000$ between trajectories that have an initial Hamming distance of 10 (0.01\% of the nodes are flipped).  The important result of these figures is that the critical stability condition $\lR = 1$ derived using our truth table annealing procedure agrees well with the numerical results from our simulations of frozen-in (quenched) systems.

\subsection{\label{subsec:strict}Networks with Strictly Canalizing Inputs}

\begin{figure}
\label{fig:fig1}
\includegraphics[scale=0.6]{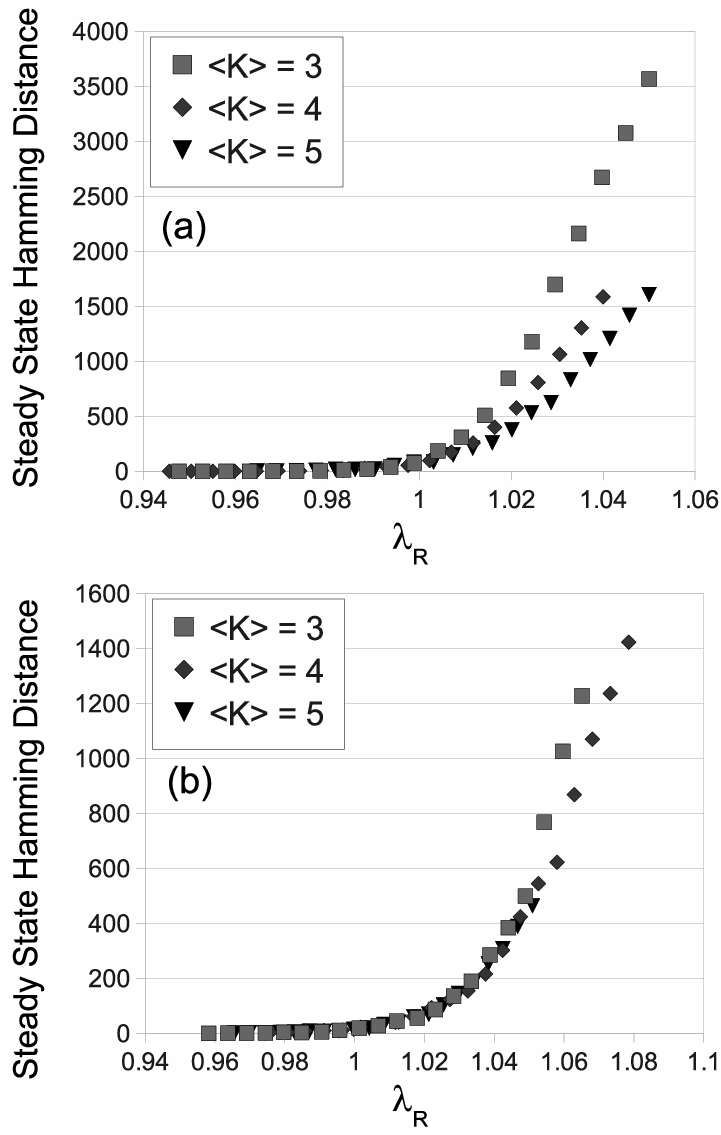}
\caption{Steady state Hamming distance vs. $\lambda_R$ for (a) $N-K$ networks with $K = 3$ (squares), 4 (diamonds), or 5 (triangles) inputs; and (b) networks with power-law in-degree distibution with average $\langle K^{in} \rangle = 3$ (squares), 4 (diamonds), or 5 (triangles).  Each datapoint consists of 1000 realizations of the truth table update functions, where the initial Hamming distance between close states is 10 nodes (0.01\%). $\lambda_R$ is varied by giving an increasing proportion of nodes a single, strictly canalizing input.  The predicted transition is at $\lambda_R = 1$.}
\end{figure}

Figure 1 is a numerical test of the theory of Sec. III on networks with an increasing proportion of nodes with a single strictly canalizing input, where the proportion increases from zero to one.  For each node in the network, we choose whether the node will have a canalizing input with probability $p_{can}$.  When $p_{can} = 0$, $\lambda_R$ takes its maximum value; when $p_{can} = 1$, $\lambda_R$ takes its minimum value.  

In order to calculate the activity with a uniform ensemble of inputs in Eq. \eqref{eq:genactivity}, some care must be exercised when constructing the truth tables for each node, and the procedure is as follows.  The canalized output for each truth table is chosen to be zero or one with equal probability, and the remaining values in the table are assigned one with probability $2p^*$ or $2p^*-1$, depending on whether the canalized output is zero or one, in order to maintain a constant truth table bias $p^*$ over all nodes.  That the canalized output is zero or one with equal probability is crucial, since we are assuming that any given set of input values to a node are equally likely in our evaluation of Eq. \eqref{eq:genactivity}.  A preponderance of canalized zeros or ones violates this assumption (although the presented calculation of the activity can be refined to take this into account as discussed in Sec. III).  Similar care must be taken in the construction of unstructured truth tables.  Since there are two solutions to $q = 2p^*(1-p^*)$ for a given $q$, the bias for each unstructured truth table is chosen to be either of the solutions with equal probability.

Figure 1(a) tests the theory on $N-K$ networks, where $K = 3, 4,$ and 5.  The networks are constructed by randomly drawing $K$ inputs to each node with uniform probability from the $N-1$ other nodes in the network, subject to the constraint that each node has exactly $K$ outputs.  The sensitivities for these networks are $q = 0.35, 0.27, $ and 0.21 for $K = 3, 4,$ and 5 networks, respectively.  According to the theory in Ref. \cite{self}, these parameters are in the disordered regime.  We see that in all cases, the networks appear to undergo a transition from ordered to disordered near $\lambda_R = 1$ as we add more canalizing inputs.  Additionally, we see that the scaling of the steady-state Hamming distance with $\lambda_R$ is a strong function of $K$, the number of inputs.  This is to be contrasted with the case of power-law degree distribution discussed below.

Figure 1(b) considers networks where the in- and out-degrees are independently drawn from truncated power-law degree distributions:  $P(K) \propto K^{-\gamma}$ if $K^{min} \leq K \leq K^{max}$, and 0 otherwise.  Networks are then constructed by randomly making connections between nodes in accord with their degrees using the configuration model.  The figure depicts  cases where the average number of inputs are the same as the $N-K$ case considered in Fig. 1(a): for $\langle K \rangle = 3$, $K^{min} = 2$, $K^{max} = 15$ and $\gamma = 2.8$; for $\langle K \rangle = 4$, $K^{min} = 2$, $K^{max} = 15$ and $\gamma = 2$; and for $\langle K \rangle = 5$, $K^{min} = 3$, $K^{max} = 15$ and $\gamma = 2.4$.  The effective sensitivities are the same as the $N-K$ cases which again place them slightly in the disordered regime.  We see again that there appears to be a transition to ordered dynamics as canalizing inputs are added.  However we note that the scaling of steady-state Hamming distance has much weaker dependence on $\langle K \rangle$ than in the $N-K$ case.  We also note that for the same $\lR$ the steady-state Hamming distance is smaller than the $N-K$ case.

%Finally, in Fig. 1(c) we consider networks with strong assortativity or disassortativity.  Networks with assortativity (disassortativity) are constructed as follows.  Starting with a power-law network constructed as above, we (cut and paste text here from my other paper).  We see (blah).

\subsection{Quasicanalizing Inputs}

\begin{figure}
\label{fig:fig2}
\includegraphics[scale=0.6]{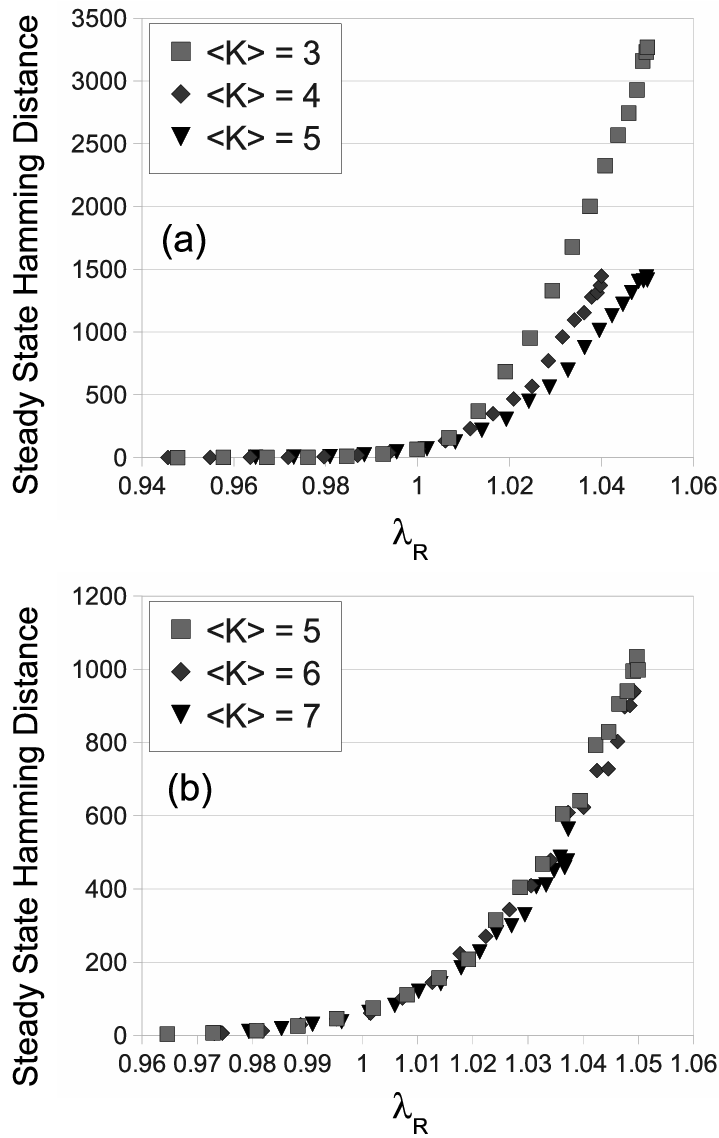}
\caption{Steady state Hamming distance vs. $\lambda_R$ for (a) $N-K$ networks with 3 (squares), 4 (diamonds), or 5 (triangles) inputs; and (b) networks with power-law in-degree distibution with average $\langle K^{in} \rangle = 3$ (squares), 4 (diamonds), or 5 (triangles).  Each datapoint consists of 1000 realizations of the truth table update functions, where the initial Hamming distance between close states is 10 nodes (0.01\%). $\lambda_R$ is varied by giving each node a single quasicanalizing input and increasing the difference  $|\gps{c}{0}-\gps{c}{1}|$, where $c$ is the index of the canalizing input.  The predicted transition is at $\lambda_R = 1$.}
\end{figure}

Figure 2 demonstrates the results of the second method of varying $\lambda_R$, where each node has a single quasicanalizing input.  The tests were done on the same network topologies as in the previous section.  When assigning a generalized canalized truth table, we randomly choose the canalizing value $v$ to be zero or one with uniform probability.   To vary $\lambda_R$, we vary $\genp{i}{c}{v}$ from zero to $\ps$: when $\genp{i}{c}{v} = 0$, all nodes have a single strictly canalizing input (i.e., the same as the case where $p_{can} = 1$ above); when $\genp{i}{c}{v} = \ps$, $c$ is no longer a canalizing input and the network is identical to the case where $p_{can} = 0$ above.  

Figure 2(a) considers $N-K$ networks where all parameters are the same as in Fig. 1(a).  We see the same behavior in steady-state Hamming distance as a function of $\lambda_R$ as in Fig. 1(a).  In Fig. 2(b), we use slightly different parameters than those in Fig. 1(b).  The parameters used in the truncated power-law distribution are as follows: for $\langle K \rangle = 5$, $K^{min} = 3$, $K^{max} = 15$ and $\gamma = 2.4$; for $\langle K \rangle = 6$, $K^{min} = 4$, $K^{max} = 15$ and $\gamma = 2.7$; and for $\langle K \rangle = 7$, $K^{min} = 5$, $K^{max} = 15$ and $\gamma = 2.9$.  
The initial sensitivities are $q = 0.21, 0.174,$ and 0.146, respectively.  These parameters were chosen in order to get larger truth tables at each node.  We note that the steady-state Hamming distance is not much larger than in the cases of Fig. 1(a), and that the scaling is an even weaker function of the average number of inputs.

\section{Effect of Correlations Between Canalization and Topology}

A useful feature of the theory of Sec. III is that it allows us to analyze the interplay between choice of canalizing input and local topology.  Here we use a perturbation approach to derive an expression for the marginal change of $\lambda_R$ as canalization is added to the network.   We will see that this result depends on the local topology at the node under consideration and the `amount' of canalization added, viz., the difference between biases when the canalizing input takes its canalizing value and when it does not.  In order to test the results in this section, we will consider networks where each node has a single strictly canalizing input, and we adjust which input is the canalizing input to vary $\lambda_R$.

In the case of no canalizing inputs, the right hand side of Eq. \eqref{eq:probf} reduces to \ps and $r_{ik} = 2\ps(1-\ps)$ for all inputs, which is the same as the unstructured case.  In the case of exactly one canalizing input and the rest non-canalizing, we get from Eq. \eqref{eq:probf}
\begin{equation}
\label{eq:activity1}
r_{ik} = \begin{cases}
\genp{i}{c}{0}(1-\genp{i}{c}{1}) + \genp{i}{c}{1}(1-\genp{i}{c}{0})  & \text{for $j = c$} \\
\genp{i}{c}{0}(1-\genp{i}{c}{0}) + \genp{i}{c}{1}(1-\genp{i}{c}{1}) & \text{for $j \neq c$},
\end{cases}
\end{equation}
where $c$ is the index of the canalizing input of node $i$.  

We now investigate the effect of adding canalizing inputs to a network.  We assume that we start with a network that has some nodes with no canalizing inputs and is characterized by an activity matrix $R$ with largest eigenvalue $\lR$.  We choose a node $i$ that is not canalized and change its truth table so that node $i$ has canalizing input $c$, but the same \ps, by perturbation analysis.  Let $R + \Delta R$ and $\lR + \Delta \lR$ denote the activity matrix and the maximum eigenvalue corresponding to the altered system. Given left and right eigenvectors $v$ and $u$ of the matrix $R$, perturbation theory gives the change in the eigenvalue as \cite{DI}
\begin{equation}
\label{eq:deltalR}
\Delta \lR \approx \frac{v^T\Delta R u}{v^T u}.
\end{equation}
Since $R_{ij} = 2\ps(1-\ps)$, by Eq. \eqref{eq:activity1} and substituting $\genp{i}{c}{1} = 2\ps - \genp{i}{c}{0}$, we see that 
\begin{equation}
\label{eq:deltaR}
\Delta R_{ij} = \begin{cases}
2(\genp{i}{c}{0})^2 + 2(\ps)^2 - 4\genp{i}{c}{0}\ps & \text{if $j = c$;} \\
-2(\genp{i}{c}{0})^2 - 2(\ps)^2 + 4\genp{i}{c}{0}\ps & \text{if $j \rightarrow i$, $j \neq c$;} \\
0 & \text{otherwise.}
\end{cases}
\end{equation}
Using \eqref{eq:deltaR} in \eqref{eq:deltalR}, we get
\begin{equation}
\label{eq:deltalR2}
\Delta \lR \approx \left( (2\genp{i}{c}{0})^2 + 2(\ps)^2 - 4\genp{i}{c}{0}\ps \right) \left(v_c u_c - \sum_{j \neq c} u_j v_j \right).
\end{equation}
The first term in the product depends on the amount of canalization; the second, however, captures the effect of the network topology on $\lR$.  We note that, since the first term is always positive, the sign of the effect on the largest eigenvalue is determined by the second term.  Usually the second term is negative, but of varying magnitude based on the choice of $c$.  In some cases, when the canalizing input has a large $u_c v_c$ compared to the other inputs, canalization may increase instability.  

\begin{figure}
\label{fig:fig3}
\includegraphics[scale=0.6]{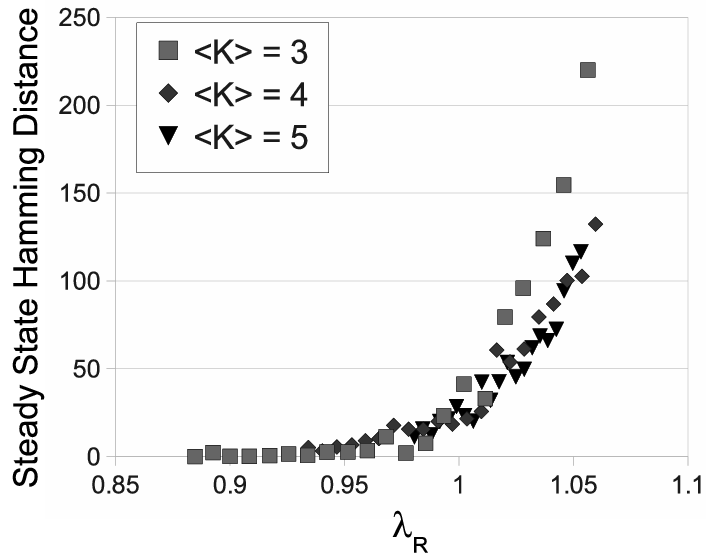}
\caption{The effect of correlation between canalization and local network topology.  Steady state Hamming distance vs. $\lambda_R$ for networks with power-law in-degree distibution with average $\langle K^{in} \rangle = 3$ (squares), 4 (diamonds), or 5 (triangles).  Each datapoint consists of 1000 realizations of the truth table update functions, where the initial Hamming distance between close states is 10 nodes (0.01\%). Each node in the network has a single, strictly canalizing input.  In this figure, $\lambda_R$ is varied by the choice of canalizing input: a varying proportion of nodes have their canalizing input chosen to maximize Eq. \eqref{eq:deltalR2}, the remainder have their canalizing input chosen to minimize Eq. \eqref{eq:deltalR2}.  The predicted transition is at $\lambda_R = 1$.}
\end{figure}

We test whether $\lambda_R$ effectively predicts the order/disorder transition in Fig. 3 using networks constructed as in Sec. IV.  All networks under consideration have $N = 10^5$ nodes and truncated power-law degree distributions corresponding to the parameters used in Fig. 1(b): for $\langle K \rangle = 3$, $K^{min} = 2$, $K^{max} = 15$ and $\gamma = 2.8$; for $\langle K \rangle = 4$, $K^{min} = 2$, $K^{max} = 15$ and $\gamma = 2$; and for $\langle K \rangle = 5$, $K^{min} = 3$, $K^{max} = 15$ and $\gamma = 2.4$. Each node in the network has a single canalizing input, and, to vary $\lambda_R$, a varying proportion of nodes $p_{max}$ have their canalizing input chosen to maximize Eq. \eqref{eq:deltalR2}, with the remaining nodes having their canalizing input chosen to minimize Eq. \eqref{eq:deltalR2}.  That is, given a list of input nodes $\{j\}$ to a node $i$, the canalizing input $c \in \{j\}$ is chosen such that $u_c v_c = \text{max}(\{u_j v_j\})$ or $u_c v_c = \text{min}(\{u_j v_j\})$, where $u_j$ is the eigenvector centrality of node $j$.  Since every node has a canalizing input,  we use slightly different values for the initial sensitivities: $q = 0.35, 0.27,$ and 0.222 for $\langle K \rangle = 3, 4, $ and 5, respectively.  Once again, as in Fig. 1(b), we see that the transition is approximately at $\lambda_R = 1$.

\section{Conclusion}

In this paper we have generalized previous work on the stability of large Boolean networks to account for canalization.  Our generalization allows a continuum in the degree of canalization and a probabilistic interpretation, as opposed to the previous canalization model \cite{PNAS1}, where an input could only be strictly canalizing or not canalizing at all.  We define a semiannealing procedure which we used to derive the condition, Eq. \eqref{eq:stability}, under which Boolean networks that have canalizing functions are stable.  The stability criterion derived in this paper offers two advantages to the study of genetic networks in particular, because it successfully handles complex network topologies that may be found in real genetic control networks, and because it can account for possible correlations between canalizing behavior and network topology.  Given the likely prominance of canalizing behavior in gene networks, these results may offer insight into the understanding of these systems.  As an example of some of the insights than can be gleaned, Ref. \cite{nykter} reports that a mutant strain of macrophage that has a gene silenced actually exhibits mildly chaotic behavior, in constrast to the unmutated case which exhibits criticality.  This may indicate that the silenced gene is a canalizing input to a large number of genes; when the canalized gene is silenced, the states of the genes it is connected to are free to evolve according to the dynamics of their other inputs, thus reducing stability and yielding a slightly chaotic network.  Furthermore, since our technique allows analysis of any specified network (e.g., an experimentally determined network), our stability criterion may, with advances in gene network measurement technology, allow one to assess the criticality of real genetic networks directly.

We thank Wolfgang Losert and the anonymous reviewers for their comments.  This work was partially supported by ONR Grant N00014-07-1-0734.

\appendix
\section{Definition of the Truth Table Draws and Constraints on $\{\gp\}$}

In order to define the appropriate annealing procedure used on the truth tables, we must specify the probability that a given output value of $f_i$ is one.  We derive this probability in two ways: first using Bayes' Theorem and again using counting arguments.  The Bayesian approach has the advantage of simplicity and clarity, however the counting approach yields the entire set of constraints on realizable sets of \gp.  In both presentations, a useful quantity is the `truth table bias' \ps, which is the probability that any truth table output is one, similar to the unstructured case.  Letting $L \equiv 2^{\Kin}$ be the number of rows in the truth table, the expected number of ones in the output column of the truth table of node $i$ is $\ps L$.  For any given arbitrary input $k$ to $f_i$, $L/2$ rows in the truth table have $s_k=0$ and $L/2$ have $s_k=1$.  By definition of \gp, the expected number of ones in the output of the truth table with entries that have $s_k=s$ is $\gp L/2$.  The total expected number of ones is the sum of the expected number of ones when $s_k=0$ and when $s_k=1$, which leads to
\begin{equation}
\label{eq:pstar}
\ps = \frac{\gps{0} + \gps{1}}{2},
\end{equation}
Note that, since the expected number of ones does not depend on our choice of $k$ above, $\gps{0} + \gps{1}$ must be independent of $k$.  This provides a constraint on the set of possible \gp values that describe a realizable truth table.  Non-canalizing inputs have both \gps{0} and \gps{1} equal to the effective bias by definition, and unstructured truth tables have $p_i = \ps$.   %Strictly canalizing inputs have \gp for the non-canalizing value equal to $2\ps$ or $2\ps-1$ depending on whether the canalized output is zero or one.  To see this, consider, for example, a truth table where the canalizing output is one, and denote the probability of an output one at node $i$ when the input is noncanalizing by $\bar{p}_i$.  Then, since the probability of both the canalizing and non-canalizing value are $1/2$, $\ps = \frac{1}{2}(1+\bar{p}_i$), yielding $\bar{p}_i = 2\ps -1$, as stated.

We now derive the probability that a given set of input values to node $i$, $\{s_1, s_2, ..., s_{\Kin}\}$, yields an output of one, and we denote this probability $\phi_i(s_1,...,s_{\Kin}) \equiv \Pr[f_i = 1 | I_1, ... I_{\Kin}]$.  Using Bayes' Theorem, we have
\begin{equation}
\label{eq:bayes1}
\phi_i(s_1, ..., s_{\Kin}) = \frac{\Pr[I_1, ... I_{\Kin}|f_i = 1] \Pr[f_i = 1]}{\Pr[I_1, ... I_{\Kin}]},
\end{equation}
where $I_k$ is the event that the $k$-th input takes the value $s_k$ (i.e., $I_k$ is the event that $\sigma_j = s_k$, where $s_k$ denotes a specifc value, 0 or 1, of the node $j(i,k)$'s state variable $\sigma_j$).  By definition, $\Pr[f_i = 1] = \ps$.  Since we are considering an ensemble where every possible input string to $f_i$ has equal probability, $\Pr[I_1, ... I_{\Kin}] = 2^{-\Kin}$.  We note that since each of the events $I_k$ are independent, $\Pr[I_1, ... I_{\Kin}|f_i = 1] = \prod_k \Pr[I_k | f_i = 1]$ and we calculate $\Pr[I_k | f_i = 1]$ again using Bayes' Theorem:
\begin{equation}
\label{eq:bayes2}
\Pr[I_k | f_i = 1] = \frac{\Pr[f_i = 1 | I_k] \Pr[I_k]}{\Pr[f_i = 1]} = \frac{\gp (1/2)}{\ps}.
\end{equation}
Using these results in Eq. \eqref{eq:bayes1}, we obtain \ref{eq:probf}.

Another method to derive \ref{eq:probf} involves a counting argument.  While this is a little less clear, it yields the full set of constraints for a realizable set of \ps.  This argument proceds from Eq. \eqref{eq:pstar}.  We now choose a second arbitrary input, $l$ and divide the truth table into four sections, each of size $L/4$: those rows with $s_k = 0, s_l = 0$; those with $s_k = 0, s_l = 1$; those with $s_k = 1, s_l = 0$; and $s_k = 1, s_l = 1$.  By definition of \gp and \genp{i}{l}{s}, we have the total expected number of ones in the truth table as
\begin{equation}
\label{eq:count2}
\frac{L}{4}\big(\genp{i}{k}{0}\genp{i}{l}{0} + \genp{i}{k}{0}\genp{i}{l}{1} + \genp{i}{k}{1}\genp{i}{l}{0} + \genp{i}{k}{1}\genp{i}{l}{1}\big) = \ps L.
\end{equation}
This yields a new set of constraints, namely that for each pair of inputs $k$ and $l$, $\genp{i}{k}{0}\genp{i}{l}{0} + \genp{i}{k}{0}\genp{i}{l}{1} + \genp{i}{k}{1}\genp{i}{l}{0} + \genp{i}{k}{1}\genp{i}{l}{1}\big) = 4\ps$.  This disallows conflicting rules, such as two strictly canalizing inputs that canalize to different values.  We can continue in a similar manner by considering a third input and deriving an equation like Eq. \eqref{eq:count2} and obtain constraints on the \gp of triplets of inputs and so on until we have divided the truth table up into L sections of one row each.

\end{document}